\documentclass[reprint, prx,  superscriptaddress]{revtex4-1}

\usepackage{verbatim}
\usepackage{graphicx}
\usepackage{animate}
\usepackage{color}
\usepackage{amsmath}
\usepackage{amssymb}
\usepackage{braket}
\usepackage{enumerate}
\usepackage{multirow}
\usepackage{colortbl}
\definecolor{Gray}{gray}{0.3}
\usepackage{wrapfig}
\usepackage[latin1]{inputenc}
\usepackage[dvipsnames]{xcolor}
\usepackage{soul}

\DeclareMathAlphabet{\mathcal}{OMS}{cmsy}{m}{n}

\newcommand{\be}{\begin{equation}}
\newcommand{\ee}{\end{equation}}

\newcommand{\mc}[1]{\mathcal{#1}}

\newcommand{\fig}[1]{Fig.~\ref{#1}}

\newcommand{\Fig}[1]{Figure~\ref{#1}}
\newcommand{\Figs}[2]{Figures~\ref{#1} and \ref{#2}}
\newcommand{\tbl}[1]{Table~\ref{#1}}
\newcommand{\eq}[1]{Eq.~\eqref{#1}}

\renewcommand{\bf}[1]{\mathbf{#1}}

\newcommand{\bea}{\begin{eqnarray}}
\newcommand{\eea}{\end{eqnarray}}
 
\newcommand{\ba}{\begin{array}}
\newcommand{\ea}{\end{array}}

\newcommand{\bl}{\begin{flalign}}
\newcommand{\enl}{\end{flalign}}

\usepackage{makecell}

\begin{document}

\title{Mapping Electronic Decoherence Pathways in Molecules}

\author{Ignacio Gustin}
\affiliation{
    Department of Chemistry, University of Rochester, Rochester, New York 14627, USA
    }
  \author{Chang Woo Kim}
\affiliation{
    Department of Chemistry, Chonnam National University, Gwangju 61186, South Korea
    }
\author{David W. McCamant}
\affiliation{
    Department of Chemistry, University of Rochester, Rochester, New York 14627, USA
    }

\author{Ignacio Franco}
\email{ignacio.franco@rochester.edu}
\affiliation{
    Department of Chemistry, University of Rochester, Rochester, New York 14627, USA
    }
\affiliation{
    Department of Physics, University of Rochester, Rochester, New York 14627, USA
    }

\date{\today}

\begin{abstract}
Establishing the fundamental chemical principles that govern molecular electronic quantum decoherence has remained an outstanding challenge. Fundamental questions such as how solvent and intramolecular vibrations or chemical functionalization contribute to the decoherence remain unanswered and are beyond the reach of state-of-the-art theoretical and experimental approaches. Here we address this challenge by developing a strategy to isolate electronic decoherence pathways for molecular chromophores immersed in condensed phase environments that enables elucidating how electronic quantum coherence is lost.  For this, we first identify resonance Raman spectroscopy as a general experimental method to reconstruct molecular spectral densities with full chemical complexity at room temperature, in solvent, and for fluorescent and non-fluorescent molecules. We then show how to quantitatively capture the decoherence dynamics from the spectral density and identify decoherence pathways by decomposing the overall coherence loss into contributions due to individual molecular vibrations and solvent modes.  We illustrate the utility of the strategy by analyzing the electronic decoherence pathways of the DNA base thymine in water. Its electronic coherences decay in $\sim 30$ fs. The early-time decoherence is determined by intramolecular vibrations while the overall decay by solvent. Chemical substitution of thymine modulates the decoherence with hydrogen-bond interactions {of the thymine ring} with  water leading to the fastest decoherence.  Increasing temperature leads to faster decoherence as it enhances the importance of solvent contributions but leaves the early-time decoherence dynamics intact. The developed strategy opens key opportunities to establish the connection between  molecular structure and quantum decoherence as needed to develop chemical strategies to rationally modulate it.
  \end{abstract}

\maketitle

Chemistry builds up from the idea that molecular structure determines the chemical and physical properties of matter. This principle guides the modern design of molecules for medicine, agriculture, and energy applications. However, chemical design principles have mostly remained elusive for emerging quantum technologies that exploit the fleeting but transformative properties of quantum coherence responsible for the wave-like interference of microscopic particles. Specifically, currently, it is not understood how chemical structure should be modified to rationally modulate quantum coherence and its loss. \cite{wasielewski2020exploiting,atzori2019second}

To leverage Chemistry's ability to build complex molecular architectures for the development of next-generation quantum technologies, there is a critical need to identify methods to tune quantum coherences in molecules and better protect them from quantum noise (or decoherence) that arises due to uncontrollable interactions of the molecular degrees of freedom of interest with its quantum environment.\cite{zhu2022functionalizing,Freedman2022,hu2022tuning,viola1999dynamical}
Protecting and manipulating molecular coherences is also necessary to unshackle chemical processes from the constraints of thermal Boltzmann statistics, as needed to enhance molecular function through coherence,\cite{scholes2017using}  develop novel routes for the quantum control of chemical dynamics,\cite{shapiro2012quantum,rice2000optical} and the design of optical spectroscopies with enhanced resolution capabilities.\cite{wang2019quantum,mukamel1995principles} 

In particular, electronic coherences generated by photoexcitation with ultrafast laser pulses in single-molecules and molecular arrays have received widespread attention.\cite{Biswas2022Coherent, dani2022time} 
These coherences dictate the photoexcited dynamics of molecules  and, as such, play a pivotal role in our elementary description of photochemistry, photophysics, charge and energy transport, and in our understanding of vital processes such as  photosynthesis and vision.\cite{de2017vibronic,nelson2018coherent,kreisbeck2012long,popp2019coherent,blau2018local,ramos2019molecular,hahn2000quantum,marsili2019two,chuang2021extreme}
These coherences decay on ultrafast (femtosecond) timescales due to the interaction of the electronic degrees of freedom with intramolecular vibrational modes and solvent.\cite{hu2018lessons,gu_generalized_2018, vacher2017electron,Arnold2017electronic, Matselyukh2022decoherence}
Despite considerable experimental and theoretical efforts to study electronic decoherence in molecules \cite{salvador2003exciton,colonna2005photon,Yang2005Photon,Biswas2022Coherent,hwang2004analysis,hwang2004electronic,shu2023decoherence,hu2018lessons,gu_generalized_2018,jasper2005electronic, Matselyukh2022decoherence}, {its} basic chemical principles are still not understood. For instance, how does electronic coherences in, say, the DNA base thymine decohere in water? How do the different vibrations in the molecule and solvent  contribute to the overall coherence loss and which one is dominant? How do chemical functionalization and isomerism influence the different contributions to the overall decoherence? Addressing these basic questions requires a general strategy to connect chemical structure of both molecule and solvent to quantum decoherence phenomena. 

From an experimental perspective, the task requires a method that enables the decomposition of the overall decoherence into individual contributions by chemical groups. This is beyond what can be done using optical absorption or fluorescence,\cite{heller1981semiclassical} photon-echo\cite{salvador2003exciton,colonna2005photon,Yang2005Photon}  and even multidimensional laser spectroscopies.\cite{Biswas2022Coherent}
To varying extents, these techniques enable characterizing the overall decoherence time scale and the molecular Hamiltonian, but do not offer means to disentangle the contributions by specific functional chemical groups. 

\begin{figure*}[htb]
    \centering
    \includegraphics[width=0.9\textwidth]{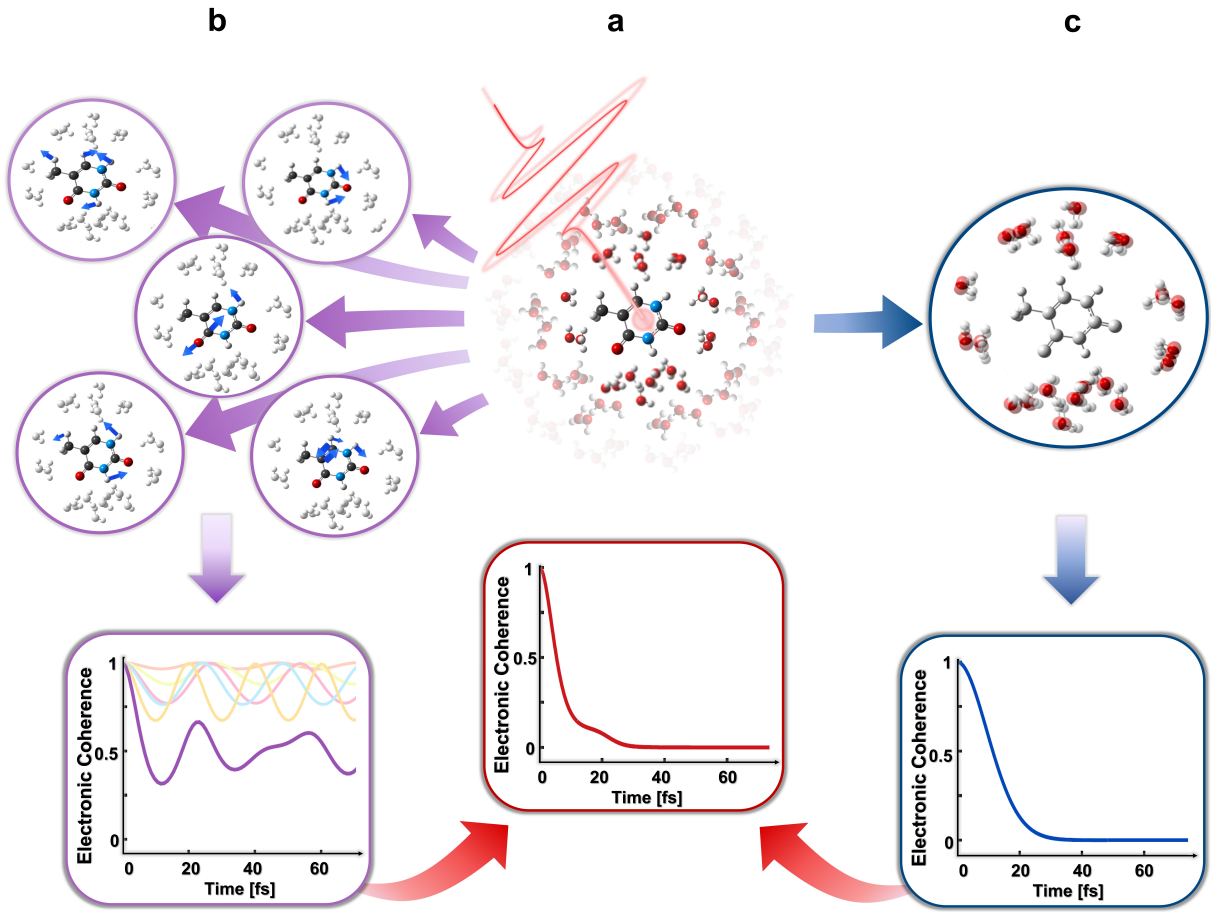}
\caption{\textbf{Mapping electronic decoherence pathways.}  (a) Photoexcitation of molecules creates electronic superposition states with coherences that decay on ultrafast timescales due to interactions of the electronic chromophore with the surrounding nuclei (solvent and intramolecular vibrations). Here we advance a method to quantify the overall decoherence and decompose it into contributions due to specific vibrational modes (b) and solvent (c), thus establishing decoherence pathways that link chemical structure with quantum decoherence. This is done by extracting the highly structured spectral densities for solvated molecules directly from resonance Raman experiments  and using the theory of decoherence functions to isolate individual contributions by solvent and vibrational modes.  The strategy opens exciting opportunities to unlock the chemical principles of electronic quantum decoherence.} 
\label{fig:Decomposition-Scheme}
\end{figure*}

From a theory perspective, the challenge is that accurately capturing the decoherence requires a fully quantum  quantitative description of the energetics and dynamics of the molecule in its chemical environment, and the computational cost of doing so increases exponentially with the size of the molecule and the environment.  For this reason, most of our theoretical understanding of decoherence arises from model problems that, while extremely useful, do not fully capture the complexities of realistic chemical systems.\cite{tomasi2005quantum,Oldbrich2011Theory} 
For instance, recent efforts to quantify electronic decoherence by explicitly propagating multidimensional quantum dynamics of molecules have so far been limited to systems at zero temperature and in vacuum,\cite{Dey2022quantum,Arnold2018control,Arnold2017electronic, hu2018lessons, Vanicek2020,Kuleff2023}
and thus are not informative of quantum decoherence  in solvent and other condensed phase environments. In turn, efforts to include the influence of thermal environments implicitly through numerically exact quantum master equations, such as the HEOM\cite{Tanimura2020HEOM,Ikeda2020Generalization} and TEMPO\cite{strathearn2018efficient}, 
have thus far  been mostly limited to simple models that do not necessarily capture the complexity of chemical environments.\cite{zhang2020proton, duan2020unusual,varvelo2023formally}

Specifically, in quantum master equations the interaction between the molecular chromophore and its thermal environment is characterized by a spectral density, $J(\omega)$\cite{Legget1987Dynamics},
which quantifies the frequencies of the nuclear environment, $\omega$, and their coupling strength with the electronic excitations. Unfortunately, the quantitative determination of $J(\omega)$ has remained elusive to theory\cite{maity2021multiscale,kim2018excited,Cignoni2022Atomistic, jang2018delocalized,Lee2016Modeling,Kell2013shape,chen2023elucidating},
and its experimental characterization is limited.\cite{ratsep2007electron,ratsep2008excitation,Pieper2009Chromophore,freiberg2009excitonic,gryliuk2014excitation}
For this reason, most studies of molecular decoherence based on quantum master equations rely on simple models of the spectral density that do not capture the  interactions of realistic chemical problems.\cite{zhang2020proton, duan2020unusual,varvelo2023formally}

Here  we introduce a general strategy to quantitatively characterize the electronic decoherence dynamics of molecules in realistic chemical environments and to map decoherence pathways, see \fig{fig:Decomposition-Scheme}. 
 The strategy is based on reconstructing the spectral density  $J(\omega)$  from resonance Raman experiments \cite{mukamel1995principles,cina2022getting}
 and decomposing the overall coherence decay into contributions by individual  vibrational and solvent modes.    Using this strategy, we can now address previously inaccessible questions, such as the interplay of solvent and vibrational contributions  and the influence of chemical substitution on quantum decoherence. The reconstructed spectral densities open opportunities to  investigate molecular decoherence dynamics in realistic chemical environments using state-of-the-art methods in quantum dynamics. Further, the overall strategy connects chemical structure to quantum decoherence dynamics and opens unique opportunities to develop the chemical principles of quantum decoherence. 

\section*{Quantum Decoherence Basics}

Quantum coherences are usually defined as the off-diagonal elements of the density matrix $\sigma(t)$ expressed in a given basis. For molecular quantum dynamics, it is useful to think about coherences in the energy basis $\sigma_{eg}(t) = \bra{e} \sigma(t) \ket{g}$ between an electronic ground $\ket{g}$ and excited state $\ket{e}$ as they lead to quantum beatings\cite{wang2019quantum} that are visible in laser spectroscopies. During decoherence electronic superposition states  with density matrix $\sigma = |\psi\rangle\langle\psi|$, where $\ket{\psi} = c_g\ket{g} + c_e\ket{e}$,  decay to a statistical mixture of states $\sigma(t) = p_g \ket{g}\bra{g} + p_e\ket{e}\bra{e}$ where $p_i\ge 0$ are the state populations. This decay is ubiquitous and  arises due to the interaction of the electrons with the surrounding nuclei (vibrations, torsions, and solvent).\cite{breuer2002theory, schlosshauer2007decoherence, joos2013decoherence} 

At zero temperature, electronic decoherence is understood as arising due to nuclear wavepacket evolution in alternative potential surfaces. 
\cite{hwang2004analysis,hwang2004electronic,prezhdo1998relationship,bittner1995quantum, jasper2005electronic,hu2018lessons, fiete2003semiclassical,shu2023decoherence}
In this view, the wavefunction of electrons and their nuclear environment is in an entangled state $\ket{\Psi(t)} = \ket{g} \ket{\chi_g (t)}+ \ket{e} \ket{\chi_e(t)}$,  where $\ket{\chi_n}$ is the nuclear wavepacket evolving in the ground $n=g$ or excited $n=e$ potential energy surface $E_n(\textbf{x})$ and  $\{\textbf{x}\}$ are the nuclear coordinates. In this case, the electronic coherences $\sigma_{eg}(t) =\bra{\chi_{e}(t)} \chi_{g}(t)\rangle$ decay with the overlap between the nuclear wavepackets in the ground and excited state. 
At finite temperatures $T$,  for early times $t$ \cite{gu_generalized_2018, gu2017quantifying, prezhdo1998relationship,bittner1995quantum} the electronic coherences  $|\sigma_{eg}(t)|^2 = |\sigma_{eg}(0)|^2 \exp{(-t^2/{\tau}_{eg}^2)}$ decay like a Gaussian with a time scale $\tau_{eg} = \frac{\hbar}{\sqrt{\langle \delta^2 {\mathcal{E}}_{eg}\rangle}}$
dictated by  fluctuations of the energy gap $\mc{E}_{eg}(\textbf{x})= E_e(\textbf{x}) - E_g(\textbf{x})$ due to thermal or quantum fluctuations of the nuclear environment. At high temperatures,
$\tau_{eg} =\frac{\hbar }{\sqrt{s k_{\mathrm{B}} T}}$  is simply connected to the Stokes shift $s$ {(the difference between positions of the maxima of absorption and fluorescence due to a given electronic transition)}.

\section*{Reconstructing Spectral Densities from Resonance Raman}

We first note that $J(\omega)$ can be reconstructed from resonance Raman (RR) spectroscopy as this technique provides detailed quantitative structural information about molecules in solution. While RR experiments are routinely used to investigate the vibrational structure and photodynamics of molecules\cite{dietze2016femtosecond,piontkowski2019excited,fang2020mapping}, 
their utility for decoherence studies had not been appreciated before. A leading technique used to reconstruct $J(\omega)$ for molecules is difference fluorescence line narrowing spectra ($\Delta$FLN).\cite{ratsep2007electron,ratsep2008excitation,Pieper2011Excitonic,Pieper2009Chromophore,freiberg2009excitonic,gryliuk2014excitation} The advantage of RR over $\Delta$FLN is that it can be used in both fluorescent and non-fluorescent molecules\cite{mccamant2004femtosecond,Mccamant2003femtosecond,kukura2007femtosecond,Lee2004Theory} (instead of only  fluorescent), in solvent (instead of glass matrices), at room temperature (instead of cryogenic temperatures), and that it offers the possibility of varying the nature of the solvent and the temperature.

In RR spectroscopy, incident light $E_{\text{L}}$ of frequency $\omega$ chosen to be at resonance with an electronic transition is scattered inelastically, yielding a Stokes and an anti-Stokes signal $E_{\text{S}}$ that changes the molecular vibrational state. {In contrast to non-resonance Raman where selection rules restrict the transitions to vibrational modes that change the molecular polarizability, in RR} all vibrational modes  that are coupled to the  electronic transition  yield a  signal.\cite{chalmers2002handbook} 
The theory of RR cross sections (see Supplementary Information and  Refs. ~\cite{Myers1982Excited,shreve1995Thermal,mukamel1995principles,Tannor1982Poly,Page1981Separation})
is based on a two-surface molecular model, with Hamiltonian $H_{M}=H_{g}\left \lvert g \right \rangle \left \langle g \right \rvert +H_{e}\left \lvert e \right \rangle \left \langle e \right \rvert$.
Here, $H_{g}=\sum_{\alpha}\left (\frac{ p^{2}_{\alpha}}{2m_{\alpha}}+\frac{1}{2}m_{\alpha}\omega^{2}_{\alpha}x^{2}_{\alpha} \right)$ is the ground-state state nuclear Hamiltonian, where $x_{\alpha}$ and $p_{\alpha}$ are the position and momentum operators of the $\alpha$-th mode of effective mass $m_{\alpha}$ and frequency $\omega_{\alpha}$. In turn, the excited potential energy surface consists of the same set of nuclear modes but displaced in conformational space, i.e., $H_{e}=\hbar \omega_{eg}+\sum_{\alpha}\left (\frac{p^{2}_{\alpha}}{2m_{\alpha}}+\frac{1}{2}m_{\alpha}\omega^{2}_{\alpha}\left(x_{\alpha}-d_{\alpha}\right)^{2} \right)$ where $\hbar \omega_{eg}$ is the electronic excitation energy. The displacement along the $\alpha$-th  mode,  $d_{\alpha}$, determines the strength of the electron-nuclear coupling as measured by the Huang-Rhys factor $S_{\alpha}= \frac{1}{2\hbar}m_{\alpha}\omega_{\alpha}d^{2}_{\alpha}$ or reorganization energy $\lambda_{\alpha}=\hbar S_{\alpha} \omega_{\alpha}$.  The overall Stokes shift $s$ is determined by  $s=2\lambda=2\sum_{\alpha}\lambda_{\alpha}$, where  $\lambda$ is the overall reorganization energy.

The RR experiments yield the frequencies of the vibrational modes  $\{\omega_\alpha\}$ and also the Huang-Rhys factors $S_\alpha$ which are extracted by fitting the signal to the RR cross section.\cite{Myers1982Excited,shreve1995Thermal,mukamel1995principles,Tannor1982Poly,Page1981Separation} The analysis also yields contributions due to solvent, and other (sub 200 cm$^{-1}$) nuclear modes that cannot be experimentally resolved, to the overall line shape and Stokes shift which are captured by an overdamped oscillator with correlation time $\frac{1}{\Lambda}$ and reorganization energy $\lambda_{0}$.\cite{li1994brownian} The procedure that is routinely done to extract these parameters is detailed in the supplementary information.

This information is exactly what is needed to reconstruct the spectral density of molecules in solvent with full chemical complexity. In this case, the spectral density consists of a broad low-frequency feature $J_{0}(\omega)$ describing the influence of solvent and a series of discrete high-frequency peaks $J_{\alpha}(\omega)$ ($\alpha=1,...,N$) due to interaction of the chromophore with intramolecular vibrational modes. The basic functional forms for the spectral density have been isolated through physical models in which the thermal environment is described as a collection of harmonic oscillators (the  so-called Brownian oscillator model).\cite{mukamel1995principles} This yields,
\be
\begin{split}
\label{eq:spectraldensity}
J(\omega)&=\sum_{\alpha=0}^{N}J_{\alpha}(\omega), \text{ where }\\
J_{\alpha}(\omega) &=\frac{2}{\pi}\lambda_{\alpha}\omega^{2}_{\alpha}\frac{\omega \gamma_{\alpha}}{\left(\omega^{2}_{\alpha}-\omega^{2}\right)^{2}+\omega^{2}\gamma^{2}_{\alpha}}
\end{split}
\ee
and  $\{\gamma_{\alpha}^{-1}\}$ are the vibrational lifetimes. The solvent spectral density can be adequately modeled via a Drude-Lorentz functional form $J_{0}(\omega)=\frac{2}{\pi}\lambda_{0}\frac{\omega\Lambda}{\omega^{2}+\Lambda^{2}}$ which arises from  $J_{\alpha}(\omega)$ when the lifetimes are short (i.e. when $\omega_\alpha/\gamma_\alpha \ll 1$).  Thus, all the parameters defining the spectral density can be extracted from the RR experiments.  While the specific value of $\gamma_{\alpha}$ cannot be directly resolved using RR,  its value plays no significant role in electronic decoherence as vibrational relaxation occurs on much longer time scales  for physical values of $\hbar\gamma_{\alpha} \in (1,20)$cm$^{-1}$ (see Fig. S1).\cite{maity2021multiscale,Bennett2013Structure,Novoderezhkin2004Energy}

The spectral densities that can be extracted from RR are due to ``pure dephasing" processes where there is no net exchange of energy between the system and the environment. That is, for processes in which the environment does not lead to electronic transitions or relaxation. In molecules, these pure dephasing processes have been the subject of intense interest as they typically occur in time scales that are much faster than the overall relaxation and dominate the decoherence dynamics. \cite{von1997laser,brinks2014ultrafast}

{Naturally, this $J(\omega)$ reconstruction is limited by experimental constraints and the scope of validity of the displaced harmonic oscillator model. For instance, internal conversion due to conical intersections occurring on time scales comparable or faster than the dephasing, or unresolvable electronic state congestion, will invalidate the analysis.  In addition, anharmonicities, frequency shifts and Duschinsky rotations (mode mixing) in the potential energy surfaces can introduce additional time scales to the decoherence that are beyond the displaced harmonic oscillator model.\cite{hwang2004electronic, hu2018lessons} 
However, numerical evidence thus far suggest that their influence occurs on time scales that are much slower than the overall dephasing in the condensed phase and thus they are expected to play a minor role. \cite{hwang2004electronic, hu2018lessons, zuehlsdorff2020nonlinear}}

\section*{Decoherence Pathways}

Once a spectral density is extracted from RR experiments, the task is then to use it to capture the realistic quantum dynamics of the molecular chromophore. The challenge is that the computational cost of doing so increases exponentially with the number of features $N+1$ in the environment. 

Second, we note that pure-dephasing processes occurring in thermal harmonic environments admit an exact solution without explicit propagation. In this case \cite{unruh1995maintaining, palma1996quantum, breuer2002theory}, the decay of the off-diagonal elements of the density matrix $|\sigma_{eg}(t)| = |\sigma_{eg}(0)|\exp[-\Gamma(t)]$ is determined by the decoherence function
\be 
\label{eq:Decoherence-function}
    \Gamma (t) =\frac{1}{\hbar}\int_{0}^{\infty}d\omega \,  J(\omega)\coth \left(\frac{\hbar\omega}{2 k_{B}T}\right)\frac{1-\cos(\omega t)}{\omega^{2}}
\ee
where $k_{B}T$ is the thermal energy. Thus,  computing the decoherence dynamics amounts to accurately calculating $\Gamma(t)$ via numerical integration  thus avoiding explicit time propagation of $\sigma(t)$.

Third, remarkably, in this context, it is possible to extract decoherence pathways in molecules by decomposing the overall
decoherence into contributions due to solvent and specific vibrational modes. Inserting  \eq{eq:spectraldensity} into \eq{eq:Decoherence-function} leads to a decomposition of the decoherence function $\Gamma(t)=\sum_{\alpha=0}^{N}\Gamma_{\alpha}(t)$ into individual contributions by specific nuclear modes
\be
\Gamma_{\alpha}(t)=\frac{1}{\hbar}\int_{0}^{\infty}d\omega  J_{\alpha}(\omega)\coth \left(\frac{\hbar\omega}{2k_{B}T}\right)\frac{1-\cos(\omega t)}{\omega^{2}}.
\ee
Alternative methods to recover the overall decoherence time such as photon-echo experiments\cite{salvador2003exciton,colonna2005photon,Yang2005Photon}, optical  absorption cross-section\cite{heller1981semiclassical} and even 2D electronic spectroscopy\cite{Biswas2022Coherent} cannot decompose the overall signal into contributions by solvent and specific vibrational modes. This unique aspect of our strategy opens a remarkable opportunity  to develop chemical insights on decoherence dynamics by isolating contributions due to particular nuclear modes.

\begin{figure}[htb]
\centering
\includegraphics[width=0.45\textwidth]{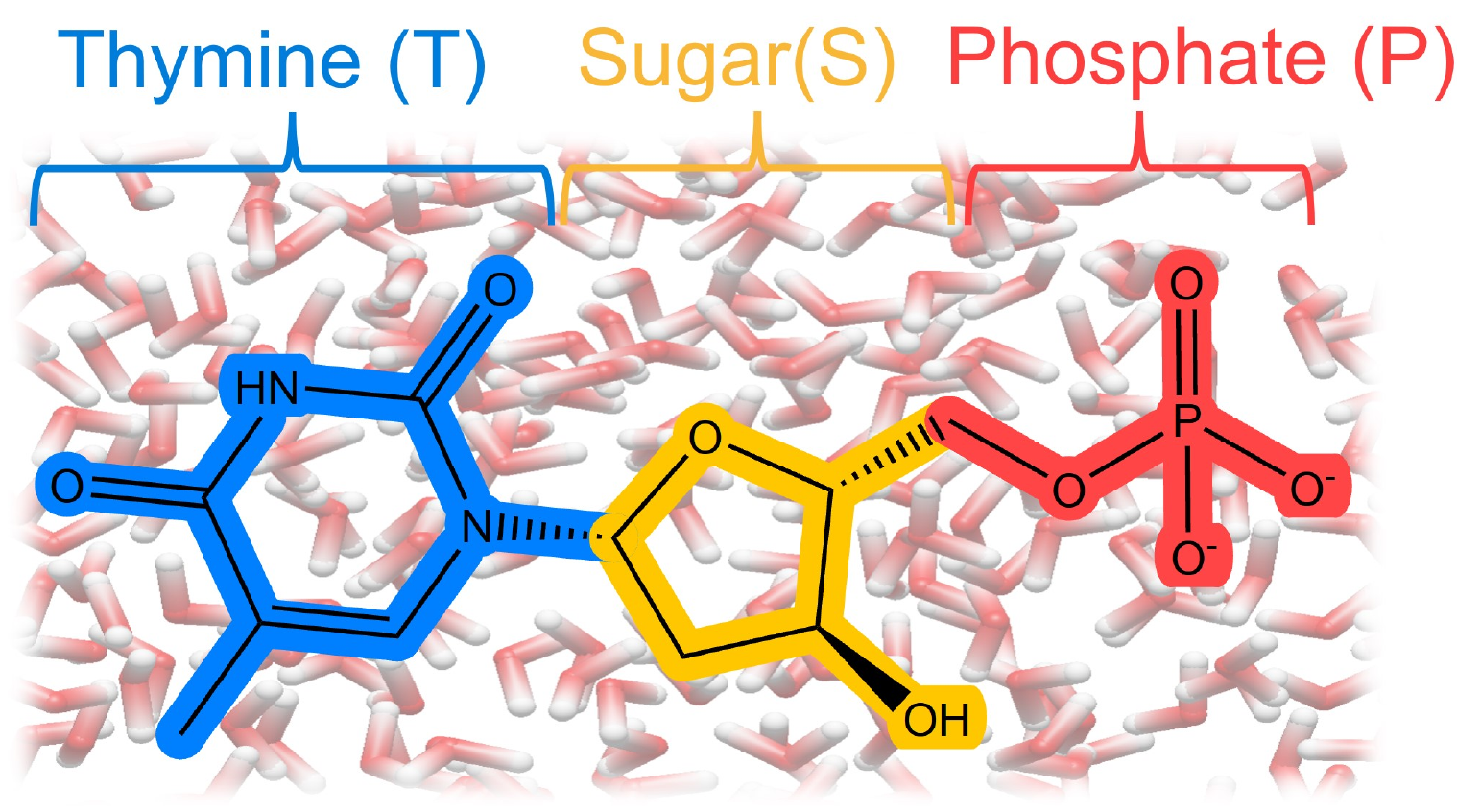}
\caption{\textbf{Molecular structure of DNA base thymine (T) its  nucleoside (T+S) and nucleotide (T+S+P)}}
\label{fig:Chemical-structures}
\end{figure}

\begin{figure}[ht]
\centering
\includegraphics[width=0.45\textwidth]{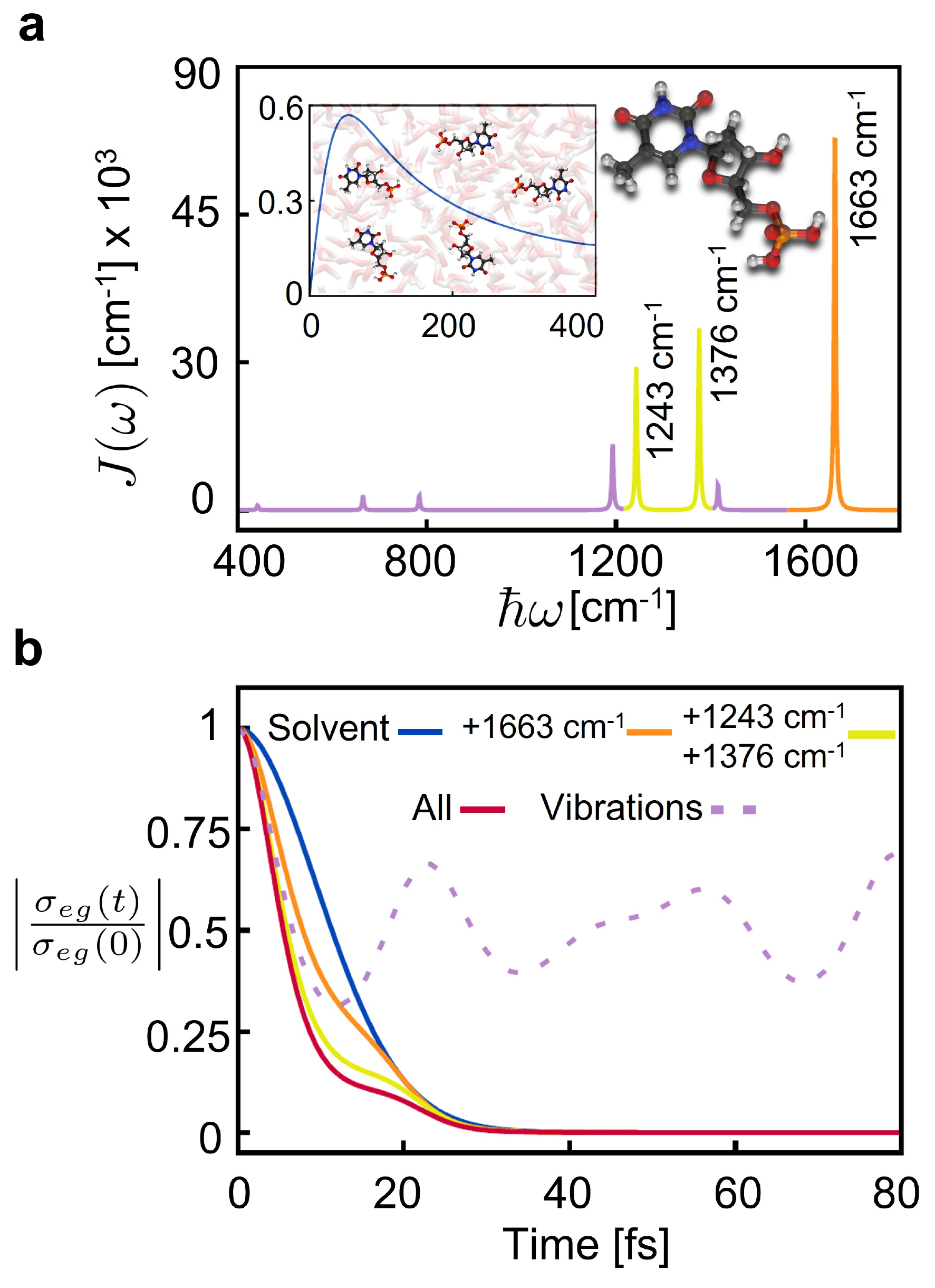}
\caption{\textbf{Decoherence dynamics of thymine nucleotide in water at 298 K.} (a) Spectral density $J(\omega)$ reconstructed from RR experiments in Ref.~\cite{yarasi2007initial, Ng2008Initial, billinghurst2012initial} with a vibrational damping factor $\hbar \gamma=4$ cm$^{-1}$. (b) Electronic decoherence (red line) and its contributions due to solvent (blue line) and intramolecular vibrations (dashed line).  The remaining curves depict decoherence due to solvent plus selected vibrational modes (orange: 1663 cm$^{-1}$; yellow: 1663 cm$^{-1}$+ 1376 cm$^{-1}$+ 1243 cm$^{-1}$).  }
\label{fig:SPD-Decoherence-Decomposition}
\end{figure}

\section*{Electronic Decoherence in a DNA nucleotide}

\subsection*{Spectral Density and Overall Decoherence}  To illustrate the power of the strategy, based on a RR analysis by Loppnow\cite{yarasi2007initial, Ng2008Initial, billinghurst2012initial} we have reconstructed the spectral density for a DNA nucleotide (thymidine 5'-monophospate), see \fig{fig:Chemical-structures}, in water at 298 K and used it to scrutinize its decoherence dynamics. 
\Fig{fig:SPD-Decoherence-Decomposition} shows the  reconstructed spectral density and the overall decoherence dynamics. The parameters used to reconstruct the spectral density are listed in Table S1 of the supplementary information.
 The spectral density (\fig{fig:SPD-Decoherence-Decomposition}a) consists of a broad low-frequency feature (blue line, inset) due to solvent and several sharp peaks due to intramolecular vibrations.  The overall decoherence (\fig{fig:SPD-Decoherence-Decomposition}b, red line) shows an initial Gaussian decay profile, followed by a partial recurrence at $\sim 20$ fs and a complete decay of coherence in $\sim 30$ fs. The Gaussian feature of the decoherence is a universal feature of initially separable system-bath states~\cite{gu_generalized_2018} and can be seen as arising due to the quantum Zeno effect. The recurrence signals the partial recovery  of the nuclear wavepacket overlap $|\bra{\chi_{e}(t)} \chi_{g}(t)\rangle|$ due to the wavepacket dynamics in the excited state potential energy surface.\cite{hu2018lessons} 
 
 {It is known that this DNA nucleotide also exhibits internal conversion due to the presence of a conical intersection. Such dynamics occurs on time scales of 100s of fs \cite{onidas2002fluorescence,gustavsson2006singlet} that are significantly longer than those required for the pure dephasing process to complete. This disparity in time scales makes the pure dephasing analysis to remain applicable even in the presence of such non-radiative electronic transitions.}

 \subsection*{Decoherence Pathways}  To understand the role of solvent and intramolecular vibrations on the decoherence, \fig{fig:SPD-Decoherence-Decomposition}b shows the contributions due to solvent (blue line), vibrations (dashed line), and solvent plus specific vibrational modes (yellow and orange lines). As shown, the early-time decoherence dynamics is dominated by vibrational contributions while the overall coherence decay is dictated by the solvent.  Out of all vibrational modes in the molecule, we find that the mode at $1663$ cm$^{-1}$,  plays the most important role. In fact, taking into account the solvent and just three {peaks due to} intramolecular vibrations  (at $1663$ cm$^{-1}$, $1376$ cm$^{-1}$ and $1243$ cm$^{-1}$) accounts for most of the decoherence dynamics (yellow line).  
 Overall, these results demonstrate the close interplay between solvent and intramolecular vibrations in the decoherence dynamics.
 

\begin{figure}[ht]
\centering
\includegraphics[width=0.45\textwidth]{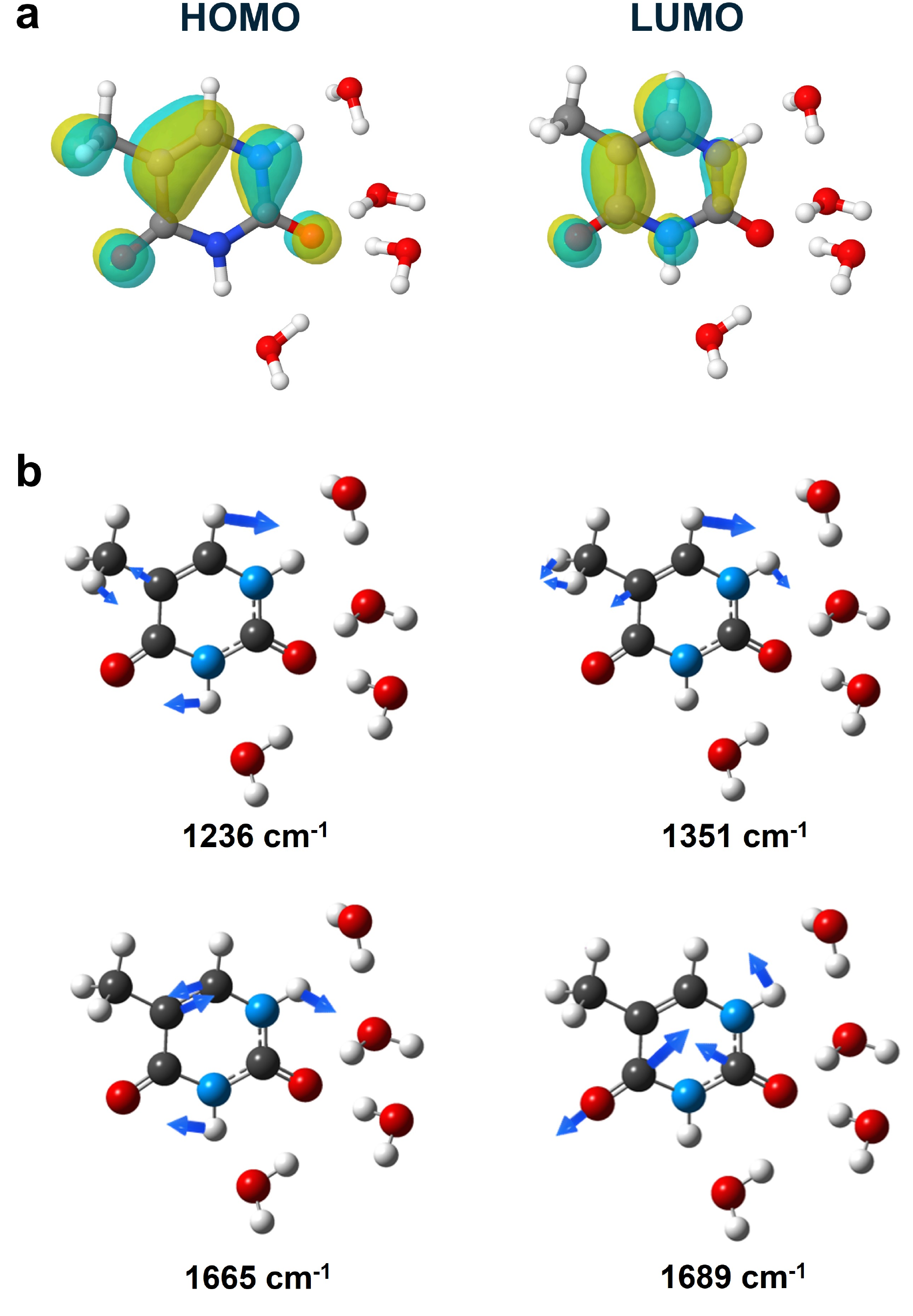}
\caption{\textbf{(a) Frontier molecular orbitals for thymine and (b) vibrational normal modes most important  for electronic decoherence.} HOMO/LUMO  stand for highest occupied/lowest unoccupied molecular orbital. The computed normal  modes  account  to the three most  prominent  peaks in the  spectral density  in \fig{fig:SPD-Decoherence-Decomposition}. The color convention is as follows: carbon (black),  hydrogen (white), nitrogen (cyan) and oxygen (red).}
\label{fig:HOMO-LUMO}
\end{figure}


\subsection*{Normal Mode Analysis}  To understand the nature of the intramolecular vibrational modes most responsible for the decoherence, we  have modeled  the resonance Raman spectra using  time-dependent density functional theory (TD-DFT)\cite{perdew1996rationale,grimme2010consistent,rappoport2010property, macak2000simulations} (computational details are included in  the Methods). To  keep the computations  tractable, we focus on  the RR spectra of  thymine  as the spectral features of  the nucleoside  and nucleotide  are nearly identical\cite{yarasi2007initial, Ng2008Initial, billinghurst2012initial} {and thus vibrational assignments in thymine are expected to also apply in the other two cases}. Quantitative agreement between theory and  experiment is obtained by including four explicit water  molecules in the first solvation shell that are an active part of the vibrational motion  of the molecule  in its microsolvated  environment and that are significantly coupled to the electronic transition.

The experimental  and  computed RR spectra are shown  in  Fig. S3. The excellent agreement between theory and experiments enables us to make vibrational assignments to  all the peaks in the spectral density. \Fig{fig:HOMO-LUMO}  details the frontier molecular  orbitals for  the electronic  transition sampled by the RR experiment and  the  4 normal modes  responsible for the most prominent peaks in the spectral density (orange/yellow  peaks in \fig{fig:SPD-Decoherence-Decomposition}a). Table S6 details the computed frequencies/Huang-Rhys factors for all 75 normal modes in the complex and Fig. S4 provides an animation  of the 10 normal modes most responsible for decoherence. The normal modes that contribute most importantly to the electronic decoherence are those that generate distortion around the thymine ring and/or the carbonyl bonds where the frontier orbitals involved in the electronic transition are located.

\begin{figure}[htb]
\centering
\includegraphics[width=0.45\textwidth]{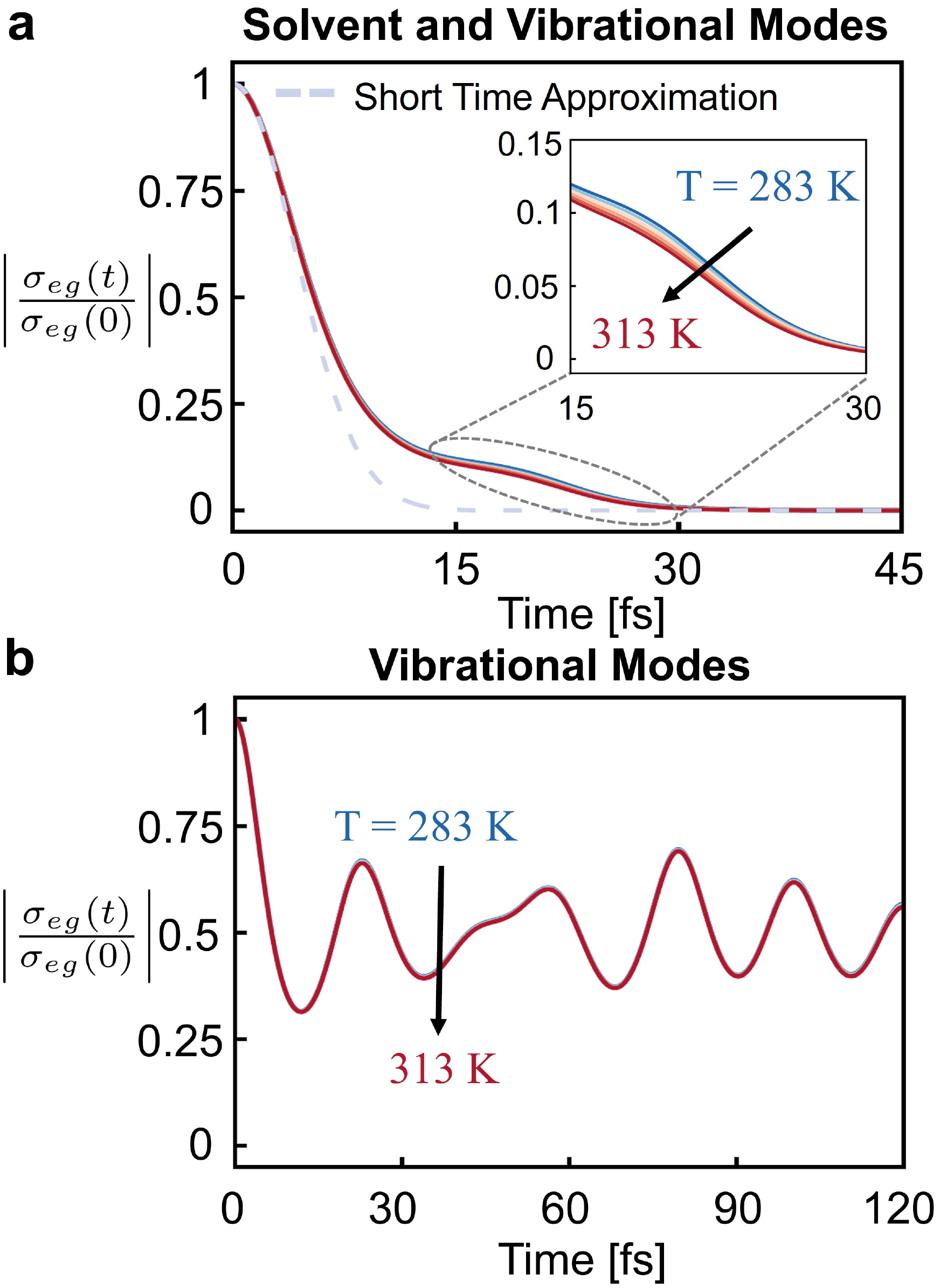}
\caption{\textbf{Temperature dependence of the electronic decoherence for thymine nucleotide in water.} (a) Full decoherence and (b) contributions due to vibrations only.}
\label{fig:Temperature}
\end{figure}

\subsection*{Temperature Dependence} To understand the role of temperature in electronic decoherence, we repeated the analysis in the {283-313 K range every 5 K}  while {taking into account that for harmonic environments $J(\omega)$ is temperature independent.}  The results in \fig{fig:Temperature} show that solvent effects become increasingly more important with temperature. In  turn,  the vibrational contributions (\fig{fig:Temperature}b) remain unaltered because thermal energy does not excite the vibrational modes that dictate decoherence. For this reason, the early-time decoherence dynamics is approximately independent of temperature as it is controlled by vibrational motion, while the overall decay is accelerated with increasing temperature leading to a reduction in the visibility of the coherence recurrence at $\sim 20$ fs. {Additional experiments are needed to isolate the importance of emergent temperature dependence of  $J(\omega)$ that can arise from differences in the mapping of chemical environments to harmonic models with varying temperature.}


\subsection*{Early Time Decoherence}
Recent efforts to understand electronic decoherence in molecules have focused on the influence of vibrational motion upon photoexcitation at 0 K. \cite{Arnold2017electronic, Arnold2018control, hu2018lessons, Dey2022quantum, Vanicek2020,Kuleff2023}
\Figs{fig:SPD-Decoherence-Decomposition}{fig:Temperature} show that this strategy accurately captures the early time coherence decay, which is dominated by high-frequency vibrations, but will not be able to correctly capture the overall coherence loss, which requires taking into account the solvent and the temperature dependence of the decoherence. 

A useful formula for electronic decoherence time scales arises by focusing on early times, which yield a Gaussian decay $|\sigma_{eg}(t)| = |\sigma_{eg}(0)|e^{-t^2/\tau_d^2}$.~\cite{prezhdo1998relationship, hwang2004analysis, gu_generalized_2018} For harmonic environments, $\tau_d^{-2} =\frac{1}{2\hbar}\int_{0}^{\infty}d\omega \,  J(\omega)\coth \left(\frac{\hbar\omega}{2k_{B}T}\right) = \frac{\hbar}{\sqrt{\langle \delta^2 {\mathcal{E}}_{eg}\rangle}}$ as can be seen by expanding the decoherence function $\Gamma(t)$ in \eq{eq:Decoherence-function} up to second order in time. \Fig{fig:Temperature}a contrasts the exact decoherence dynamics with this Gaussian form (dashed line). As shown, this simple estimate accurately captures the early time decay but does not capture possible recurrences and overestimates the overall decoherence by a factor of $\sim 2$. Thus, such formulas are useful to understand the early stages of decoherence but exaggerate the decoherence rate for electrons in molecules.

Nevertheless, to capture the electronic decoherence for molecules immersed in solvent and  other condensed phase  environments, \fig{fig:Temperature} underscores the utility of the simple Gaussian estimate as it captures the  early-time  decoherence dynamics correctly while avoiding explicitly solving the quantum dynamics. By contrast, computationally expensive efforts to capture the decoherence based on propagating the correlated  quantum dynamics of electrons and nuclei at  0 K using  Multi Configurational Time-Dependent Hartree (MCTDH)~\cite{Manthe1992Wave,beck2000multiconfiguration, Arnold2017electronic, Arnold2018control, Dey2022quantum, Kuleff2023} are only relevant for early times where finite  temperature effects due to solvent do not play a role. \Fig{fig:Temperature} shows that this early-time region  of the decoherence can be accurately captured while avoiding propagating the quantum dynamics all together. 

\subsection*{Effect of Chemical Substitution}
\begin{figure}[htb]
\centering
\includegraphics[width=0.45\textwidth]{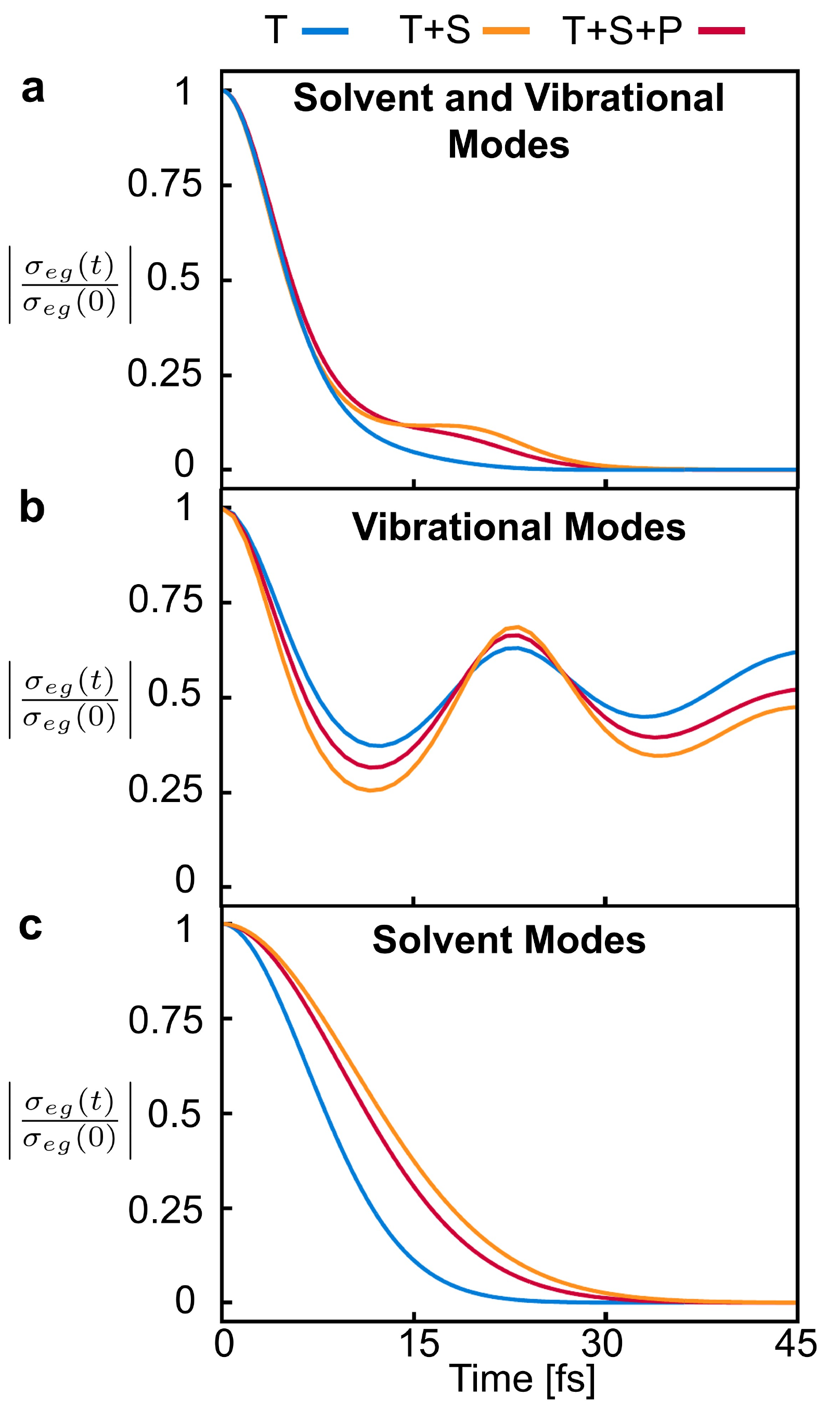}
\caption{\textbf{Effect of chemical substitution on electronic decoherence.} (a) Full electronic decoherence for the series (See \fig{fig:Chemical-structures}). (b) Electronic decoherence due to vibrations only. {(c)  Electronic decoherence due to solvent only}. Spectral densities were reconstructed from data in Ref. \cite{billinghurst2012initial,yarasi2007initial}.}
\label{fig:Chemical-subs}
\end{figure}

How does varying chemical structure influences quantum coherence loss? To address this question, we reconstructed the spectral densities of thymine (T), its nucleoside (T + sugar (S)) and nucleotide (T + S + phosphate (P)) in water at 298 K from RR data in Refs. \cite{billinghurst2012initial,yarasi2007initial}.  \Fig{fig:Chemical-subs} details the overall decoherence and the contributions of intramolecular vibrations {and solvent} to the decoherence in this series{, and \fig{fig:Chemical-structures} their chemical structure}. The reconstructed $J(\omega)$ for thymine and its  nucleoside, as well as all the needed parameters, are included in Fig. S2 and Table S2-S3 of the supplementary information, respectively.

{ Surprisingly,  the overall decoherence in thymine is the fastest, even when it is the smallest and least polar molecule in the set and admits fewer possible hydrogen-bond interactions with the solvent.  This can be rationalized by noting that the electronic excitation is localized in the thymine ring (see \fig{fig:HOMO-LUMO}) and thus interactions with the solvent with atoms in this area are expected to have the strongest impact on the decoherence.  In fact, in T the N-H termination in the ring, which is absent in T+S and T+S+P, strongly interacts with water through hydrogen bonding\cite{beyere2004dependence}, accelerating the overall decoherence.}

Coherence recurrences are prominent in T+S because solvent-induced decoherence is the slowest in this case {(see \Fig{fig:Chemical-subs}c)}. By contrast, T shows no recurrences due to ultrafast decoherence due to the solvent.
\Fig{fig:Chemical-subs}b details the decoherence due to intramolecular vibrations. The trend for the early time decoherence due to vibrations is opposite to the overall behavior with T showing the slowest decoherence, followed by T+S+P and then T+S. Thus, the overall decoherence trends in \Fig{fig:Chemical-subs}a are due to both contributions of solvent and intramolecular vibrations, even at early times.


 \section*{Conclusions}
In conclusion, we have advanced a strategy that can be used to establish the chemical principles that underlie electronic quantum coherence loss in molecules.  Previously inaccessible fundamental questions such as how chemical functionalization or solvent character contribute to the overall electronic decoherence, or how to  develop chemical strategies to  rationally modulate the decoherence can now be systematically addressed. 
The strategy is based  on reconstructing spectral densities from  resonance Raman experiments on molecular chromophores in solvent, and using the theory of decoherence functions to decompose the overall decoherence dynamics into contributions by individual solvent and vibrational modes, thus establishing decoherence pathways in molecules.  

Using this strategy we were able to scrutinize, for the first time, the electronic decoherence dynamics of the DNA base thymine in room temperature water. In this case, the decoherence occurs in $\sim 20-30$ fs with the early stages dictated by intramolecular vibrations, while the overall decoherence time due to interactions with solvent. Chemical substitution of the thymine with sugar and sugar phosphate reveals a recurrence in the  coherence dynamics that is suppressed  by hydrogen bonding between thymine and water.  The vibrational modes most responsible for the decoherence are  those that distort the thymine ring and carbonyl bonds. Increasing the temperature accelerates the contributions of solvent to the decoherence but leaves the early-time decoherence dynamics intact. For pure-dephasing, this early stage  of the decoherence is accurately captured by the theory of decoherence time scales \cite{gu_generalized_2018, gu2017quantifying, prezhdo1998relationship,bittner1995quantum} which does not require propagating the quantum dynamics.

We envision the application of this strategy to elucidate the role of diverse chemical and biological environments to the decoherence, and to establish decoherence  pathways for localized and delocalized, spin-specific and charge-transfer electronic excitations,  and in molecules  of  varying size and rigidity.  Further, the extracted spectral densities will be of general utility to characterize the dynamics of molecular chromophores with full chemical complexity using quantum master equations  \cite{Tanimura2020HEOM,Ikeda2020Generalization, strathearn2018efficient}  of increasing sophistication, {and to advance our computational capabilities to accurately capture electron-nuclear interactions in condensed phase environments.\cite{chen2023elucidating,maity2021multiscale,Cignoni2022Atomistic,kim2018excited}}.

\section*{Materials and Methods}

\subsection*{RR spectra from TD-DFT} To compute the resonance Raman spectra for thymine we adopted the following  computational strategy. First, to determine the solvation  structure  around  thymine we used  Autosolvate to create  microsolvated cluster structures.\cite{hruska2022autosolvate} 
Using a two-layer ONIOM,  we performed a  preliminary analysis of the normal modes of  the microsolvated thymine.  In the ONIOM analysis\cite{dapprich1999new,chung2015oniom} 
the molecule was treated at a quantum mechanical (QM) level while the  solvent was captured through  a  classical force  field. For the QM computation we used Density Functional Theory (DFT) with the PBE0 functional and 6-311+G(d,p) as a basis. The normal mode analysis revealed four water molecules that most strongly couple to the molecule. The geometry of the thymine and the 4 water molecules were then optimized using DFT using both PBE0/6-311+G(d,p) and CAM-B3LYP/6-311+G(d,p). The overall dielectric environment was captured through a self-consistent reaction field (SCRF) \cite{tomasi2005quantum} as the polarizable continuum model. The XYZ coordinates of the optimized geometries are shown in Table S4 and S5.

Using these structures, the resonance Raman spectra for thymine was computed using the vertical gradient\cite{macak2000simulations} approach. Ground state potential energy surfaces were determined through DFT while electronic transitions and excited-state gradients through time-dependent DFT.\cite{perdew1996rationale,grimme2010consistent,rappoport2010property} The solvent was captured through the SCRF and the four explicit water molecules. The spectra was  computed using both PBE0/6-311+G(d,p) and CAM-B3LYP/6-311+G(d,p).  All computations were done using the Gaussian 16 simulation  package.\cite{frisch2016gaussian} The computed frequencies and Huang-Rhys factors for the 75 normal modes in the complex are included in Table S6.  Figure S3 compares the  computed RR spectra with the experimental results in Refs.~\cite{yarasi2007initial, Ng2008Initial}

\subsection*{Integration of the Decoherence Function.} To accurately compute the integral in  \eq{eq:Decoherence-function} we employed the trapezoidal rule with a frequency step of 0.6 cm$^{-1}$. Recognizing the integrand as even, we extended the integration to both negative and positive frequencies. This approach effectively limited the error in solvent reorganization energy to below 0.1$\%$ and in total vibrational reorganization energy to under 0.001$\%$.

\begin{acknowledgments}
    This work was supported by a Pump Primer II research award of the University of Rochester. I.F. is supported by the National Science Foundation under Grant  No. CHE-2102386.
\end{acknowledgments}
\clearpage
\onecolumngrid
\setcounter{figure}{0} 
\renewcommand{\thefigure}{S\arabic{figure}}
\renewcommand{\thetable}{\arabic{table}}
\section*{Supplementary Information}
\section{Extracting the Spectral Density from Resonance Raman Spectra}
For completeness, below we include standard expressions for the main cross sections that result from the theory of absorption,  fluorescence and Raman scattering. We also sketch how the quantities involved are extracted in practice from experiments. This procedure is needed to extract the spectral density from resonance Raman experiments. 

The cross-sections as a function of frequency $\omega$ for absorption, fluorescence and Raman scattering are given by:\cite{Page1981Separation,shreve1995Thermal,Tannor1982Poly,Myers1982Excited} 
\begin{eqnarray}
\label{eq:A-cross-section}
    \sigma_{A}(\omega) & =& \frac{4\pi e^{2}\omega}{3c\eta\hbar}\int d \omega_{eg} G(\omega_{eg}) \text{Im}\left \{i \int_{0}^{\infty} dt \:e^{it(\omega-\omega_{eg})-g(t)}A(t)\right \}; \\
\label{eq:F-cross-section}
    \sigma_{F}(\omega)&=& \frac{4\pi\eta e^{2}\omega}{3c\hbar}\int d \omega_{eg} G(\omega_{eg}) \text{Im}\left \{i \int_{0}^{\infty} dt \:e^{it(\omega-\omega_{eg})-g^{*}(t)}A^{*}(t)\right \};\\
\label{eq:RR-cross-section}
\sigma_{R}(\omega,\omega_{S})&=& \frac{8e^{4}\omega^{3}_{S}\omega}{9c^{4}\hbar^{2}}\int d \omega_{eg} G(\omega_{eg}) \times\\
&  &  \text{Im}\left \{ 
    i
    \int_{0}^{\infty} d\tau e^{i\tau(\omega-\omega_{eg})}d(\tau)
    \int_{0}^{\infty} dt_{1} 
    e^{it_{1}(\omega-\omega_{eg})-g(t_{1})}
    \int_{0}^{\infty} dt_{2} 
 e^{-it_{2}(\omega-\omega_{eg})-g^{*}(t_2)} F(t_{1},t_{2},\tau)
    \right \}. \nonumber
    \end{eqnarray}
Here, $\omega_{eg}$ is the 0-0 transition energy between the ground and excited electronic states, $\eta$ the refractive index, and $\omega_{S}$ the frequency of the scattered radiation during the Raman process. The shape of the absorption, emission, and Raman  spectral features  are determined by the solvent-induced electronic line broadening function,
\begin{equation}
    g(t)=\frac{1}{\hbar} \int_{0}^{\infty} J_{0}(\omega) \left [\coth{\left(\frac{\beta \hbar \omega}{2}\right)}\frac{1-\cos{(\omega t)}}{\omega^{2}}+i\frac{\sin{(\omega t)}-\omega t}{\omega^{2}} \right]d\omega.
\end{equation}
Here $\beta=\frac{1}{k_{B}T}$ is the inverse temperature and $J_{0}(\omega)$ is the solvent spectral density. The quantity $J_{0}(\omega)$  is well-described by a  Drude-Lorentz functional form (see main text) which, in turn, is fully determined by the solvent correlation time $\frac{1}{\Lambda}$ and reorganization energy $\lambda_{0}$. The Raman cross section has an additional vibrational lineshape $d(\tau)$ which is usually described through a normalized Lorentzian.

The effect of intramolecular electron-nuclear interactions on the cross-sections is captured by the absorption $A(t)$ and resonance Raman $F(t_{1},t_{2},\tau)$, time correlators: 
\begin{subequations}\label{correlators}
    \begin{align}
        A(t) &= \sum_{i} P_{i} \sum_{\rho} \left \langle i \right \rvert e^{iH_{g}t/\hbar}\mu_{\rho}^{*}e^{-iH_{e}t/\hbar}\mu_{\rho}\left \lvert i \right \rangle \\
        F(t_{1},t_{2},\tau)&=\sum_{i} P_{i} \sum_{\rho\xi} \left \langle i \right \rvert  e^{-iH_{g}t_{2}/\hbar}\mu_{\xi}^{*}  e^{iH_{e}t_{2}/\hbar}\mu_{\rho} e^{-iH_{g}\tau/\hbar}\mu_{\rho}^{*} e^{-iH_{e}t_{1}/\hbar}\mu_{\xi} e^{iH_{g}t_{1}/\hbar}e^{iH_{g}\tau/\hbar} \left \lvert i \right \rangle.
    \end{align}
\end{subequations}
Here $\mu$ is the electronic transition dipole moment along the $\rho$ or $\xi$ cartesian axis, $P_{i}$ is the Boltzmann thermal occupation  of vibrational state $|i\rangle$, and  $H_{g}$ and $H_{e}$ are the ground and excited state vibrational Hamiltonians. In turn, the inhomogeneous broadening function, $G(\omega_{eg})$, is 
expressed as a normalized  distribution of 0-0 energies around an average energy $\bar{\omega}_{eg}$ 
\begin{equation}
    G(\omega_{eg})= \frac{1}{\theta \sqrt{2\pi}} e^{-\frac{(\bar{\omega}_{eg}-\omega_{eg})^{2}}{2\theta^{2}}}
\end{equation}
where $\theta$ is the standard deviation of the  distribution. 

The complexity of evaluating these cross sections is that they require quantum dynamical propagation of vibrational wavepackets along  multidimensional ground and electronic potential energy surfaces. To make further progress, it is useful to specialize considerations to the displaced harmonic oscillator model 
where the ground and excited state potential energy surfaces are described by $N$-dimensional harmonic surfaces with identical nuclear frequencies $\{\omega_\alpha\}$ and masses $\{m_\alpha\}$ that are  displaced in conformational space, with displacement $d_\alpha$ along normal mode $\alpha$. As shown by Shreve (Eqs. 15 and 17 in Ref. \cite{shreve1995Thermal}), for this model it is possible to analytically solve the quantum dynamics in Eq. \ref{correlators} and provide formulas for the cross sections that depend on the normal mode frequencies $\omega_{\alpha}$, the Huang-Rhys factor $S_{\alpha}=\frac{1}{2\hbar}m_{\alpha}\omega_{\alpha}d_{\alpha}^{2}$ or reorganization energy $\lambda_{\alpha}=\hbar S_{\alpha}\omega_{\alpha}$, and the remaining spectroscopic parameters that define the cross sections. 
The total reorganization energy of the model is given by the sum of intramolecular and solvent reorganization energy $\lambda=\lambda_{0}+\sum_{\alpha}\lambda_{\alpha}$.

The absorption, emission, and resonance Raman excitation profiles can be effectively modeled using equations Eqs. \ref{eq:A-cross-section}-\ref{eq:RR-cross-section}. The cross section units are  $\mathring{\text{A}}^{2}$/molecule, enabling direct comparison with experiment.
Adjusting the model's parameters, \{$\mu$, $\omega_{eg}$, $\theta$, $\omega_{\alpha}$, $S_{\alpha}$, $\lambda_{0}$, $\Lambda$\}, allows the experimental profiles to be accurately fitted.
Specifically, the temperature and refractive index are fixed by the experimental conditions. Initial values for the transition dipoles can be estimated from results for similar molecules and the 0-0 transition frequency, $\omega_{eg}$, can be approximated by the crossing point between absorption and fluorescence spectra. The solvent parameters, $\lambda_{0}$ and $\frac{1}{\Lambda}$, are related by the ratio
\begin{equation}
    \kappa = \sqrt{\frac{\hbar \Lambda^{2}}{2\lambda_{0}k_{B}T}},
\end{equation}
with $\kappa$ found to be 0.1 for a polar solute in polar solvents\cite{mukamel1995principles,li1994brownian,Yan1987Molecular}. Solvent parameters and the full width at half maximum of
the optical absorption lineshape, $\Gamma$, are closely linked by\cite{mukamel1995principles}
\begin{equation}
    \Gamma= \sqrt{\frac{2\lambda_{0}k_{B}T}{\hbar}}\frac{2.355+1.76 \kappa}{1+0.85\kappa+0.88\kappa^{2}}.
\end{equation}
The initial $\Gamma$ value can be estimated from the behavior of similar molecules.
 The vibrational frequencies, $\{\omega_{\alpha}\}$, are obtained from the experimental resonance Raman spectra and the Huang-Rhys factors are initially estimated under the approximation $S_{\alpha}=\frac{I_{\alpha}}{2 \omega_{\alpha}}$ where $I_{\alpha}$ is the Raman integrated peak intensity.
 
 In the fitting procedure, the parameters $\Gamma$, $\mu$, and $\omega_{eg}$ are optimized first as they have a large and uniform effect on all three lineshapes. 
 Then, the set $\{S_{\alpha}\}$ is adjusted to obtain the best fit to the absorption, emission, and resonance Raman excitation profiles, while accurately reproducing the experimental Stokes shift. The inhomogeneous broadening parameter is neglected in the fitting procedure and is only introduced when a better fit cannot be achieved. The parameters obtained from the fitting procedure can be utilized to reconstruct the spectral density, employing Eq. 1 in the main text.
\clearpage

\section{Effect of vibrational broadening on decoherence}

\begin{figure}[htb!]
\centering
\includegraphics[width=0.7\textwidth]{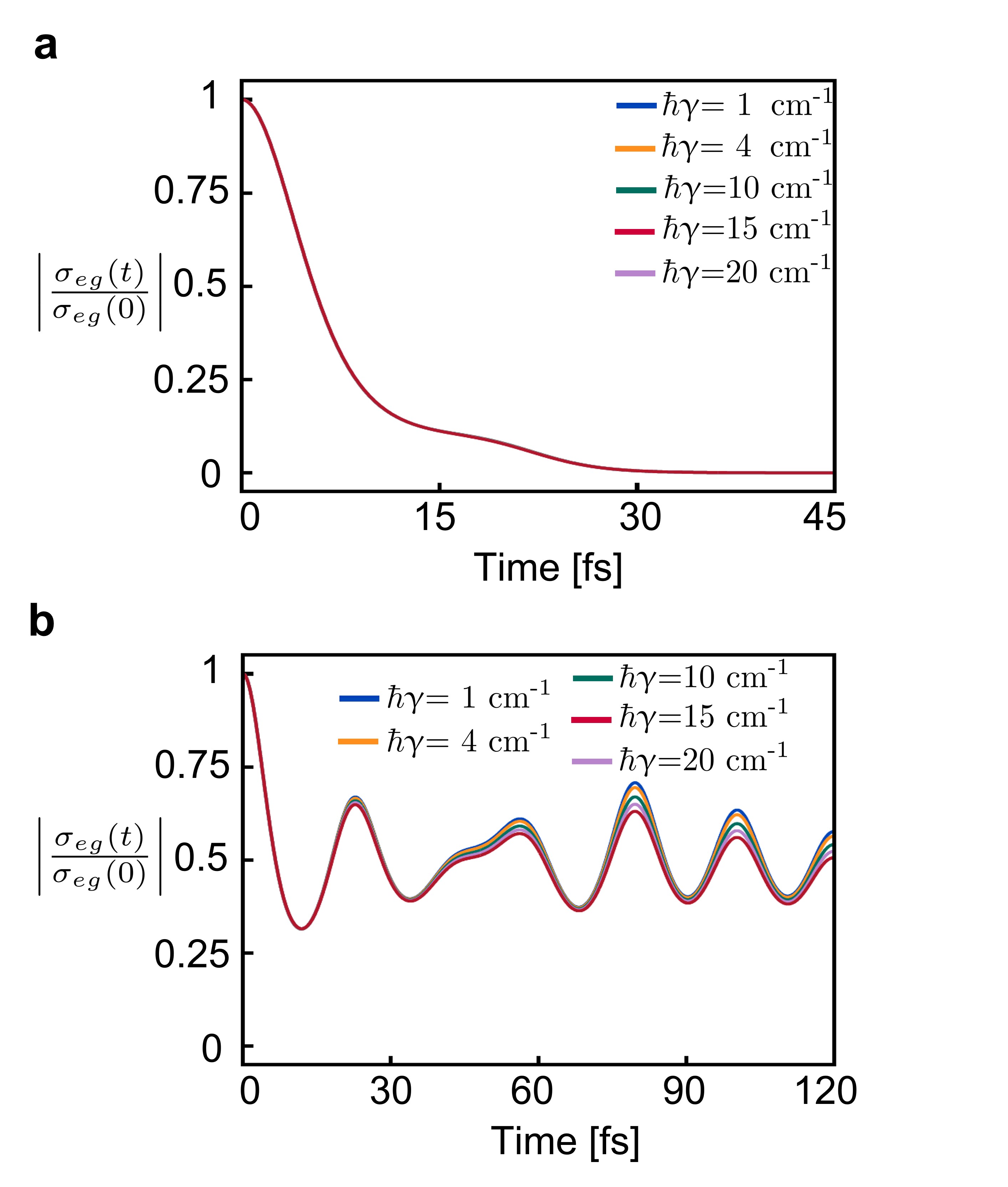}
\caption{\textbf{Effect of the vibrational broadening factor $\gamma$ on electronic decoherence}.  Coherence loss due to (a) solvent + vibrations and   (b) vibrations only for varying $\gamma$. As shown, the vibrational broadening in physically meaningful ranges has a negligible effect on electronic coherence loss, as its influence only becomes important at longer time scales. }
\label{fig:broadening}
\end{figure}

Figure \ref{fig:broadening} shows the influence of changing the vibrational broadening $\gamma$ on the electronic coherence loss. As shown, for the physically accessible range,   $\gamma$ has a negligible influence  on electronic decoherence.

\clearpage

\section{Extracted spectral densities for thymine and its derivatives}
\begin{figure}[htb!]
\centering
\includegraphics[width=0.7\textwidth]{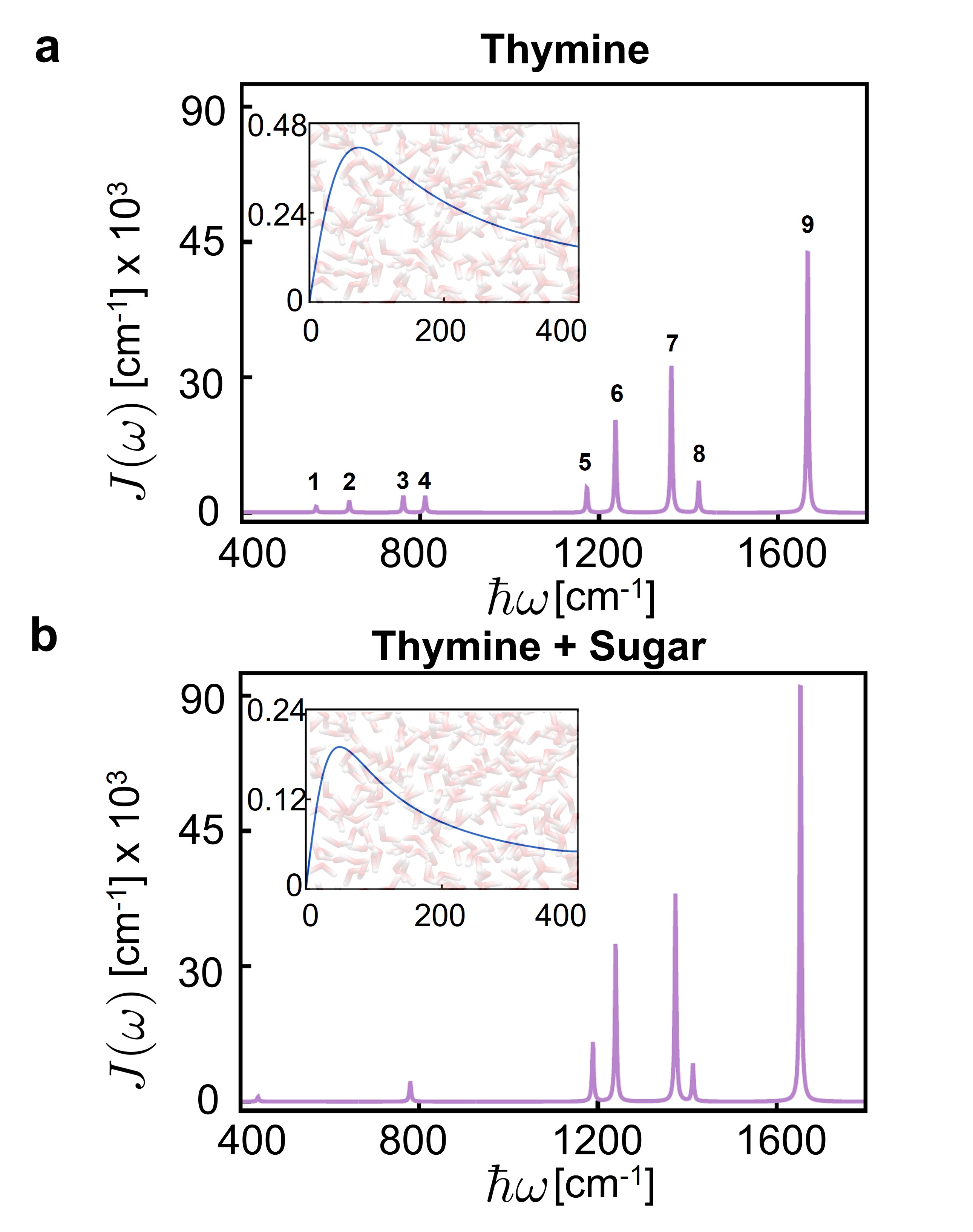}
\caption{\textbf{Reconstructed Spectral Densities} for (a) thymine and (b) its nucleoside in water at 298 K.  Labels 1-9 correspond to the normal modes in \fig{fig:normalmodes}. }
\label{fig:SPD}
\end{figure}

Tables \ref{Reorganization-Energies1}-\ref{Reorganization-Energies3} summarize the  intramolecular and solvent parameters needed for reconstructing the spectral density  for all three molecules in Fig. 2. Figure \ref{fig:SPD} details the spectral density  for thymine and its nucleoside.

\begin{table}[htb!] 
\begin{center}
\caption{Spectral density parameters for thymine nucleotide.  Solvent parameters: $\lambda_{0}=715.7\text{ cm}^{-1}$, $\Lambda=54.5\text{ cm}^{-1}$. Data obtained from Ref. \cite{billinghurst2012initial} }\label{Reorganization-Energies3}%
\begin{tabular}{@{} |c|c |c|@{}}
\toprule
Mode Frequency [cm$^{-1}$]  &  Huang-Rhys factor $S$  &  Reorganization energy $\lambda$  [cm$^{-1}$]\\
\hline
 442   & 0.034    & 14.9\\
 665   & 0.048     & 31.9\\
 784   & 0.034    & 26.5\\
 1193   & 0.065     & 77.3\\
 1243   & 0.13    & 161.6\\
 1376   & 0.135     & 186\\
 1416   & 0.018    & 25.6\\
  1663   & 0.198    & 330\\
\toprule
\end{tabular}
\end{center}
\end{table}

\begin{table}[htb!] 
\begin{center}
\caption{Spectral density parameters for thymine nucleoside.  Solvent parameters: $\lambda_{0}=596.2\text{ cm}^{-1}$, $\Lambda=49.7\text{ cm}^{-1}$. Data obtained from Ref. \cite{billinghurst2012initial} }\label{Reorganization-Energies2}%
\begin{tabular}{@{} |c|c |c|@{}}
\toprule
Mode Frequency [cm$^{-1}$]  &  Huang-Rhys factor $S$  &  Reorganization energy $\lambda$  [cm$^{-1}$]\\
\hline
 439   & 0.036    & 16\\
 780   & 0.051     & 39.9\\
 1189   & 0.065    & 77\\
 1240   & 0.162     & 201.4\\
 1374   & 0.168    & 231.1\\
 1413   & 0.029     & 40.7\\
 1645   & 0.245    & 405.2\\
\toprule
\end{tabular}
\end{center}
\end{table}

\begin{table}[htb!] 
\begin{center}
\caption{Spectral density parameters for thymine.  Solvent parameters: $\lambda_{0}=1302.9\text{ cm}^{-1}$, $\Lambda=73.5\text{ cm}^{-1}$. Data obtained from Ref. \cite{yarasi2007initial} }\label{Reorganization-Energies1}%
\begin{tabular}{@{} |c|c |c|@{}}
\toprule
Mode Frequency [cm$^{-1}$]  &  Huang-Rhys factor $S$  &  Reorganization energy $\lambda$  [cm$^{-1}$]\\
\hline
 567   & 0.028    & 16.3\\
 641   & 0.045     & 28.8\\
 762   & 0.045    & 34.3\\
 811   & 0.039     & 31.8\\
 1173   & 0.029    & 33.8\\
 1237   & 0.097     & 119.7\\
 1362   & 0.125    & 170.2\\
 1423   & 0.024     & 34.4\\
 1667   & 0.151    & 252.1\\
\toprule
\end{tabular}
\end{center}
\end{table}

\clearpage

\section{Computed resonance Raman spectra for thymine}

The resonance Raman spectra for thymine was computed as described in the Materials and Methods section.  The XYZ coordinates of the optimized geometries of the complexes used in the computations are shown in Table \ref{coordinate1} and Table \ref{coordinate2}.  The computed frequencies and Huang-Rhys factors for the 75 normal modes in the complex are included in Table \ref{normalmodedata}.   Figure \ref{fig:RR} compares the  computed RR spectra with the experimental results in Refs.~\cite{yarasi2007initial, Ng2008Initial} The excellent  agreement between theory and experiment  enables us to assign specific vibrational modes to the peaks in the spectral density and interpret the electronic decoherence pathways.  Animated Figure \ref{fig:normalmodes} details representative normal modes 1-9 that most strongly  couples to the electronic transition. 

\begin{figure}[htb!]
\centering
\includegraphics[width=0.7\textwidth]{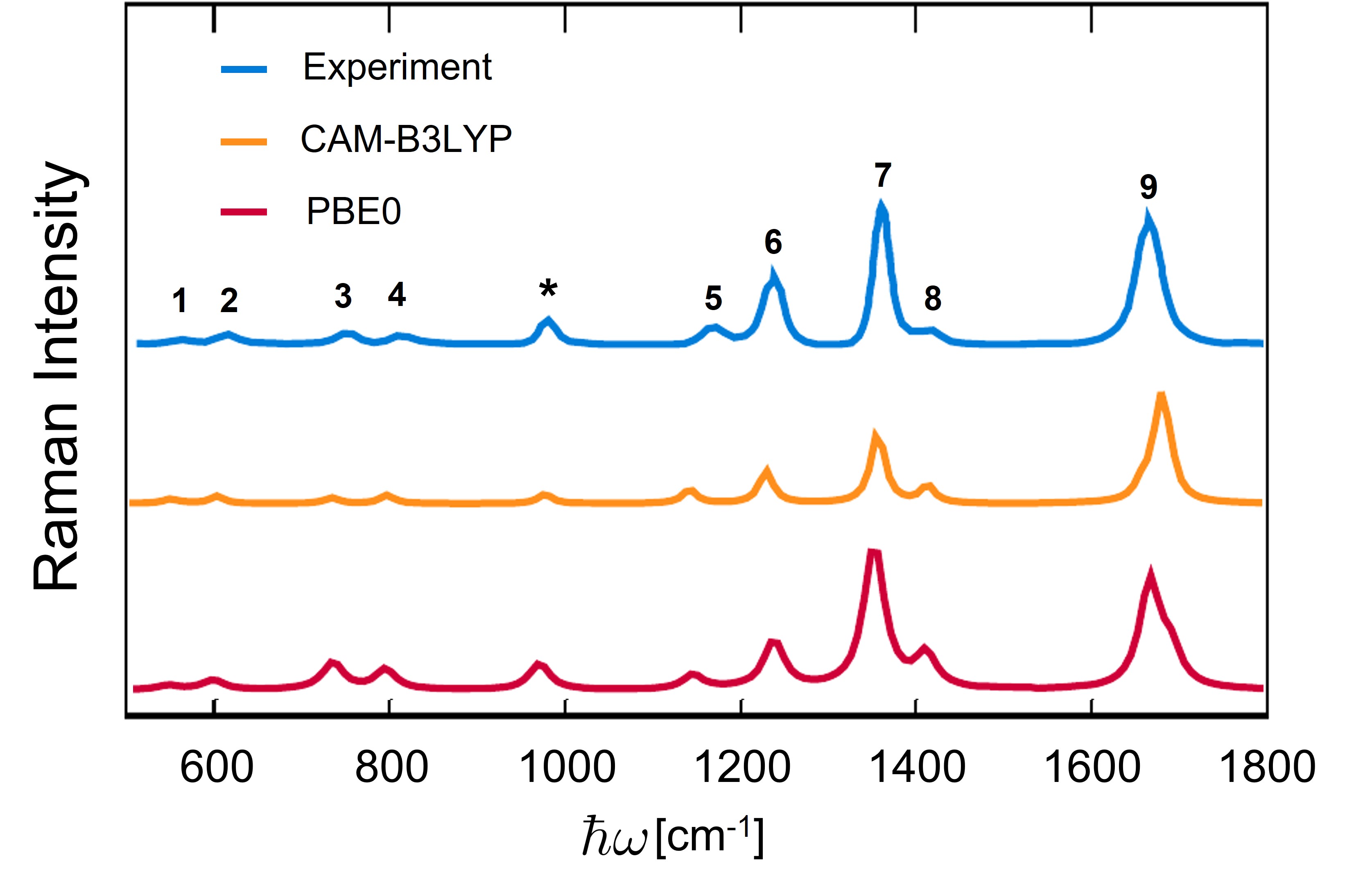}
\caption{\textbf{Comparison between calculated and experimental RR spectra for thymine.\cite{billinghurst2012initial}} The Raman excitation wavelength is 266 nm, and the functional frequencies have been scaled according to literature scaling factors for CAM-B3LYP (0.9613) and PBE0 (0.9512).\cite{merrick2007evaluation} The quantitative agreements allows us to make vibrational assignments. The asterisk marks the experimental internal standard.  Labels 1-9 correspond to the normal modes in Fig. \ref{fig:normalmodes}. The experimental RR spectra was adapted with permission from  \cite{billinghurst2012initial}. Copyright 2012 American Chemical Society  }
\label{fig:RR}
\end{figure}

\begin{figure}[htb!]
\centering
\includegraphics[width=0.7\textwidth]{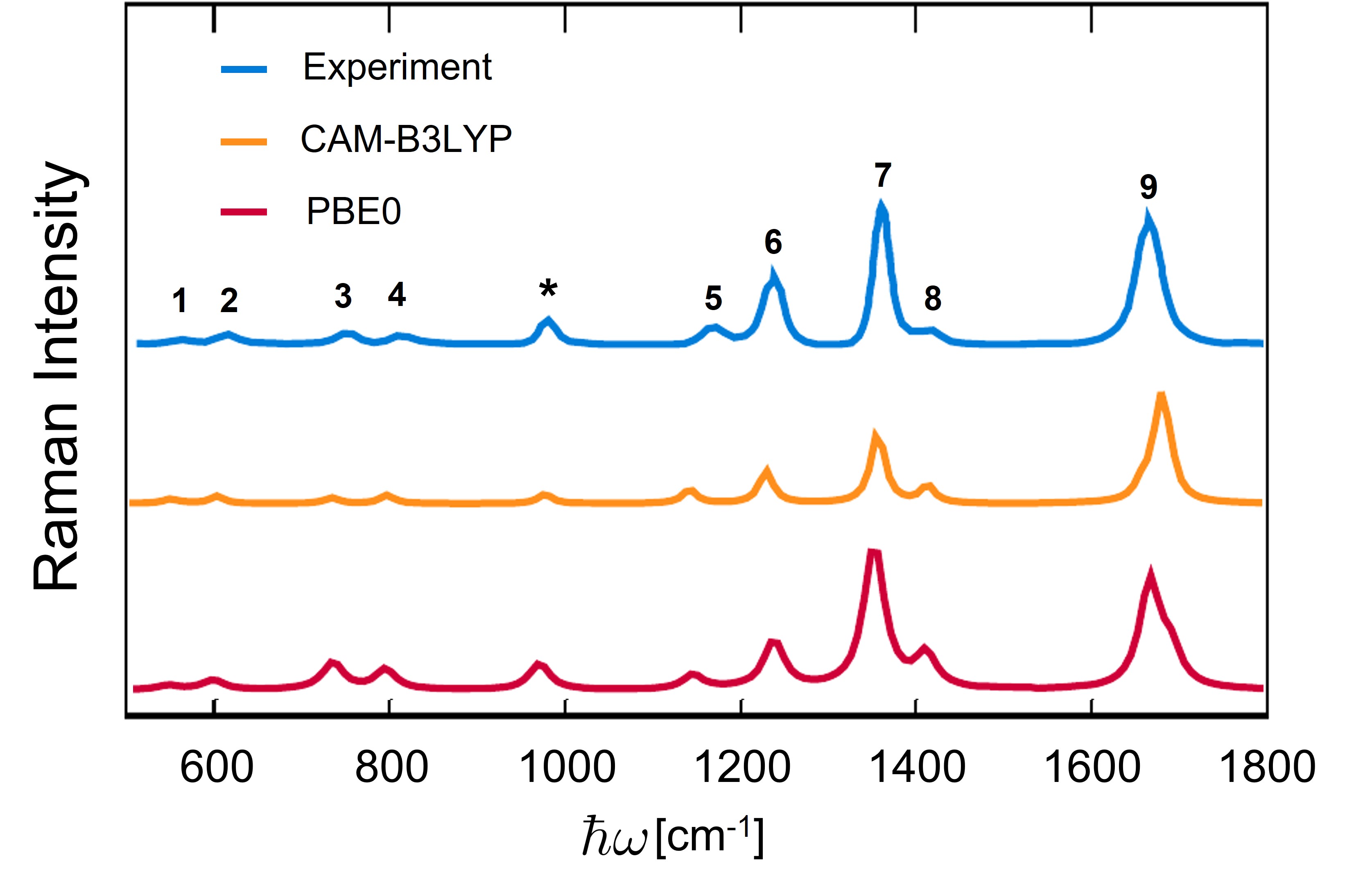}
\caption{\textbf{Comparison between calculated and experimental RR spectra for thymine.\cite{billinghurst2012initial}} The Raman excitation wavelength is 266 nm, and the functional frequencies have been scaled according to literature scaling factors for CAM-B3LYP (0.9613) and PBE0 (0.9512).\cite{Ng2008Initial,merrick2007evaluation} The quantitative agreements allows us to make vibrational assignments. The asterisk marks the experimental internal standard.  Labels 1-9 correspond to the normal modes in \fig{fig:normalmodes}.}
\label{fig:RR}
\end{figure}

\begin{table}[htb!] 
\begin{center}
\caption{Cartesian coordinates of thymine-water complex with ground-state energy optimized using PBE0/6-311+G(d,p) level of theory with SCRF.}\label{coordinate1}%
\begin{tabular}{@{} |c|c |c|c|@{}}
\toprule
Atom   & X & Y & Z \\
\hline
C  &     -2.298864 &  -0.360076 &  -0.319497 \\
N  &     -1.003836 &  -0.862288 &  -0.354389 \\
C  &      0.108847 &  -0.267634 &   0.138846 \\
N  &     -0.093939 &   0.937181 &   0.716834 \\
C  &     -2.439529 &   0.936629 &   0.319065 \\
C  &     -1.330471 &   1.521347 &   0.805037 \\
C  &     -3.794800 &   1.548191 &   0.399962 \\
H  &     -4.477269 &   0.902513 &   0.953392 \\
H  &     -4.216033 &   1.685889 &  -0.596278 \\
H  &     -3.752455 &   2.515516 &   0.896321 \\
H  &     -0.849747 &  -1.768425 &  -0.800394 \\
H  &      0.712647 &   1.402454 &   1.099369 \\
H  &     -1.345439 &   2.484616 &   1.293254 \\
O  &      1.225135 &  -0.793560 &   0.063601 \\
O  &     -3.211347 &  -1.001981 &  -0.804257 \\
O  &      0.378590 &  -3.081638 &  -1.434347 \\
H  &      0.438850 &  -3.989602 &  -1.141002 \\
H  &      1.072148 &  -2.608714 &  -0.964139 \\
O  &      3.494884 &   0.580578 &  -0.829242 \\
H  &      4.255647 &   0.229532 &  -0.369024 \\
H  &      2.738422 &   0.076340 &  -0.498377 \\
O  &      3.102473 &   3.325965 &  -0.766230 \\
H  &      3.069676 &   3.600623 &  -1.680091 \\
H  &      3.253787 &   2.368987 &  -0.800202 \\
O  &      2.627371 &  -1.943346 &   2.277198 \\
H  &      2.126894 &  -1.572948 &   1.542294 \\
H  &      2.249337 &  -2.809914 &   2.413494 \\ 
\toprule
\end{tabular}
\end{center}
\end{table}

\begin{table}[htb!] 
\begin{center}
\caption{Cartesian coordinates of thymine-water complex with ground-state energy optimized using CAM-B3LYP/6-311+G(d,p) level of theory with SCRF.}\label{coordinate2}%
\begin{tabular}{@{} |c|c |c|c|@{}}
\toprule
Atom   & X & Y & Z \\
\hline
C & -2.326166 & -0.341637 & -0.374253 \\
N & -1.026666 & -0.849412 & -0.386018 \\
C & 0.066204 & -0.281992 & 0.188660 \\
N & -0.161152 & 0.889441 & 0.828755 \\
C & -2.492730 & 0.921563 & 0.333224 \\
C & -1.407144 & 1.473656 & 0.897765 \\
C & -3.856109 & 1.532827 & 0.391440 \\
H & -4.563462 & 0.857952 & 0.877734 \\
H & -4.235904 & 1.730011 & -0.613122 \\
H & -3.835763 & 2.471166 & 0.944968 \\
H & -0.857326 & -1.733726 & -0.875809 \\
H & 0.632447 & 1.333719 & 1.269400 \\
H & -1.440905 & 2.408178 & 1.440566 \\
O & 1.190447 & -0.805181 & 0.127573 \\
O & -3.219883 & -0.953566 & -0.932662 \\
O & 0.446328 & -3.009241 & -1.502304 \\
H & 0.527831 & -3.940438 & -1.273746 \\
H & 1.092445 & -2.534325 & -0.957143 \\
O & 3.497294 & 0.521983 & -0.764408 \\
H & 4.253624 & 0.153379 & -0.297437 \\
H & 2.715598 & 0.046707 & -0.429374 \\
O & 3.294370 & 3.280833 & -0.924304 \\
H & 3.349177 & 3.501582 & -1.858136 \\
H & 3.374030 & 2.308120 & -0.881284 \\
O & 2.647056 & -1.837874 & 2.357708 \\
H & 2.120987 & -1.512185 & 1.610298 \\
H & 2.432709 & -2.772485 & 2.430093 \\
\toprule
\end{tabular}
\end{center}
\end{table}

\begin{table}[htb!] 
\begin{center}
\caption{Thymine normal modes frequencies and Huang-Rhys factors from DFT. The Huang-Rhys factors have been normalized. The labels 1-9 in bold correspond to the normal modes in \fig{fig:normalmodes}.}\label{normalmodedata}%
\begin{tabular}{@{} |c|c |c|c|@{}}
\toprule
Mode Frequency PBEO [cm$^{-1}$]   & Huang-Rhys factor S & Mode Frequency CAM-B3LYP [cm$^{-1}$] & Huang-Rhys factor S \\
\hline
11.33 &  0.009 & 14.69 &  0.070\\
21.45 &  0.033 & 17.54 &  0.022\\
27.94 &  0.019 & 27.98 &  0.000\\
29.68 &  0.005 & 32.16 &  0.029\\
37.49 &  0.125 & 46.62 &  0.000\\
40.91 &  0.010 & 57.97 &  0.002\\
50.65 &  0.002 & 66.80 &  0.001\\
57.32 &  0.001 & 70.14 &  0.000\\
71.66 &  0.002 & 93.04 &  0.011\\
81.86 &  0.048 & 115.81 & 0.000\\
110.83 &  0.001 & 150.61 & 0.269\\
142.85 &  0.250 & 161.80 & 0.001\\
151.09 &  0.012 & 166.14 & 0.001\\
162.96 &  0.085 & 175.67 & 0.176\\
164.29 &  0.274 & 196.45 & 0.003\\
192.00 &  0.003 & 206.37 & 0.000\\
206.53 &  0.008 & 222.49 & 0.035\\
224.06 &  0.082 & 251.54 & 0.008\\
233.29 &  0.001 & 294.13 & 0.045\\
284.48 &  0.034 & 304.16 & 0.000\\
291.67 &  0.002 & 315.72 & 0.005\\
301.15 &  0.002 & 336.17 & 0.003\\
316.45 &  0.003 & 388.64 & 0.001\\
385.21 &  0.010 & 419.98 & 0.266\\
\textbf{401.08 (1)}  &  \textbf{0.237} & 420.80 & 0.007\\
414.55 &  0.000 & 442.93 & 0.001\\
419.05 &  0.006 & 499.92 & 0.192\\
482.19 &  0.099 & 556.47 & 0.000\\
512.30 &  0.001 & 558.91 & 0.009\\
527.44 &  0.000 & 592.24 & 0.194\\
\textbf{572.02 (2)} &  \textbf{0.167} & 611.15 & 0.022\\
596.37 &  0.012 & 647.23 & 0.000\\
616.10 &  0.000 & 649.01 & 0.286\\
\textbf{626.73 (3)} &  \textbf{0.288} & 773.30 & 0.006\\
747.26 &  0.013 & 789.86 & 0.167\\
761.55 &  0.000 & 807.95 & 0.000\\
\textbf{770.65 (4)} &  \textbf{0.460} & 815.04 & 0.000\\
782.26 &  0.000 & 857.53 & 0.223\\
\textbf{833.47 (5)} &  \textbf{0.284} & 877.88 & 0.000\\
848.87 &  0.000 & 981.67 & 0.000\\
943.41 &  0.000 & 1051.36 &  0.188\\
1018.46 &  0.233 & 1096.85 &  0.004\\
1065.56 &  0.002 & 1122.24 &  0.000\\
1078.05 &  0.000 & 1228.23 &  0.228\\
1202.78 &  0.106 & 1279.11 &  0.000\\
1256.67 &  0.005 & 1321.97 &  0.484\\
\textbf{1299.59 (6)} &  \textbf{0.320} & 1459.14 &  0.960\\
\textbf{1419.79 (7)} &  \textbf{0.974} & 1486.39 &  0.001\\
1435.58 &  0.012 & 1519.69 &  0.225\\
1480.09 &  0.000 & 1528.88 &  0.000\\
\textbf{1481.75 (8)} &  \textbf{0.235} & 1544.22 &  0.001\\
1500.46 &  0.001 & 1553.01 &  0.000\\
1507.75 &  0.003 & 1590.18 &  0.000\\
1559.88 &  0.002 & 1654.80 &  0.005\\
1629.17 &  0.000 & 1675.82 &  0.000\\
1650.36 &  0.000 & 1681.49 &  0.000\\
1657.42 &  0.003 & 1708.83 &  0.002\\
1678.83 &  0.000 & 1783.09 &  0.183\\
\textbf{1750.56 (9a)} &  \textbf{1.000} & 1804.90 &  1.000\\
1752.42 &  0.046 & 1815.27 &  0.377\\
\textbf{1775.69 (9b)} &  \textbf{0.346} & 3183.84 &  0.002\\
\toprule
\end{tabular}
\end{center}
\end{table}

\begin{table}[htb!] 
\begin{center}

\begin{tabular}{@{} |c|c |c|c|@{}}
\toprule
Mode Frequency PBEO [cm$^{-1}$]   & Huang-Rhys factor S & Mode Frequency CAM-B3LYP [cm$^{-1}$] & Huang-Rhys factor S \\
\hline
3099.66 &  0.005 & 3243.54 &  0.000\\
3166.16 &  0.000 & 3273.18 &  0.000\\
3195.53 &  0.000 & 3372.26 &  0.009\\
3282.88 &  0.007 & 3524.17 &  0.004\\
3432.97 &  0.000 & 3703.70 &  0.000\\
3661.04 &  0.009 & 3779.26 &  0.000\\
3700.04 &  0.006 & 3783.55 &  0.002\\
3726.16 &  0.000 & 3853.91 &  0.000\\
3802.59 &  0.004 & 3884.43 &  0.000\\
3840.61 &  0.008 & 4047.99 &  0.000\\
3982.62 &  0.002 & 4055.70 &  0.000\\
3991.37 &  0.001 & 4057.12 &  0.000\\
3992.55 &  0.003 & 4057.27 &  0.000\\
3993.70 &  0.005 & &\\
\toprule
\end{tabular}
\end{center}
\end{table}

\begin{frame}{}
  \begin{figure}[htb!]
    \centering
    \animategraphics[loop,autoplay,width=\linewidth]{10}{Gif/something-}{0}{23}
    \caption{\textbf{Frontier molecular orbitals for thymine and  vibrational normal  modes that couple most strongly with the electronic transition.} HOMO/LUMO  stand for highest occupied/lowest unoccupied molecular orbital.  The labels 1-9 for the normal modes correspond to peaks 1-9 in the  spectral density (\fig{fig:SPD}) and the resonance Raman spectra (\fig{fig:RR}). The normal modes frequencies and Huang-Rhys factors  are mark in bold in \tbl{normalmodedata} . The color convention is as follows: carbon (black),  hydrogen (white), nitrogen (cyan) and oxygen (red). This is an animated figure that  shows  the  normal mode motions and is best seen in Acrobat or Okular.  \label{fig:normalmodes}}
  \end{figure}
\end{frame}
\twocolumngrid
\clearpage
\bibliography{sn-bibliography}
\end{document}